\documentstyle[11pt,aaspp4]{article}

\def\today{\ifcase\month\or
  January\or February\or March\or April\or May\or June\or
  July\or August\or September\or October\or November\or December\fi
  \space\number\day, \number\year}

\slugcomment{ApJ, in press}
\received{September 7, 1998}
\accepted{May 10, 1999}
\journalid{ApJ}{}
\articleid{00}{00}
\lefthead{Kashyap \& Drake}
\righthead{X-ray Variability on Active Binaries}

\begin{document}

\title{On X-ray Variability in Active Binary Stars}
\author{Vinay Kashyap\altaffilmark{1} and Jeremy J.\ Drake\altaffilmark{2}}

\affil{$^1$Harvard-Smithsonian Center for Astrophysics, 
MS-83, \\ 60 Garden Street, \\ Cambridge, MA 02138}
\affil{$^2$Harvard-Smithsonian Center for Astrophysics, 
MS-3, \\ 60 Garden Street, \\ Cambridge, MA 02138}

\authoremail{kashyap@cfa.harvard.edu}
\authoremail{jdrake@cfa.harvard.edu}

\begin{abstract}

We have compared the X-ray emissions of active binary stars observed
at various epochs
by the {\it Einstein} and {\it ROSAT} satellites in order to investigate the
nature of their X-ray variability.  The main aim of this work is to
determine whether or not active binaries exhibit long-term variations in
X-ray emission, perhaps analogous to the observed cyclic behavior of solar
magnetic activity.  We find that, while the mean level of emission of the
sample remains steady, comparison of different {\it ROSAT}
observations of the same stars shows significant variation on timescales
$\lesssim 2$ yr, with an ``effective variability'' $\frac{\Delta I}{I} =
0.32 \pm 0.04$,
where $I$ and $\Delta I$ represent the mean, and variation from the mean,
emission, respectively. A comparison of {\it ROSAT} All-Sky Survey and
later pointed observations with earlier observations of the same stars
carried out with {\it Einstein} yields only marginal evidence for a larger
variation ($\frac{\Delta I}{I} = 0.38 \pm 0.04$
for {\it Einstein} vs. {\it ROSAT} All-Sky Survey and $0.46 \pm 0.05$
for {\it Einstein} vs. {\it ROSAT} pointed) at these longer timescales
($\sim 10$ yr), indicating the possible presence of a long-term 
component to the variability.

Whether this long-term component is due to the presence of cyclic
variability cannot be decided on the basis of existing data.  However,
assuming that this component is analogous to the observed cyclic
variability of the Sun, we find that the relative magnitude of the
cyclic component in the {\it ROSAT} passband can at most be a factor
of 4, i.e., $\frac{I_{cyc}}{I_{min}} < 4$ .  This is to be compared to
the corresponding -- significantly higher -- solar value of
$\sim 10 - 10^2$ derived from GOES, Yohkoh, and Solrad data.  These 
results are consistent with the suggestions of earlier studies that a 
turbulent or distributive dynamo might be responsible for the 
observed magnetic activity on the most active, rapidly rotating stars.

\end{abstract}


\section{Introduction}

Observations of the solar corona over timescales of years have shown
the coronal X-ray emission, together with other indicators of activity
such as Ca\,II H and K emission line strength, to be modulated by the
solar dynamo on the 22 year magnetic field polarity reversal cycle,
with maxima and minima occuring every 11 years or so (e.g., see the
review by Harvey 1992).  Surveys of the X-ray sky performed by the {\it
Einstein Observatory}, and later by {\it EXOSAT} and {\it ROSAT}, have
also firmly established the existence of supposedly analogous hot
X-ray emitting coronae throughout the late main sequence (F-M), and
also in late-type giants down to spectral types near mid-K (e.g.,
Vaiana et al.\ 1981).  One fundamental issue in stellar physics
concerns the relationship between this magnetic activity on stars with
a wide range of physical parameters and solar magnetic activity (see
review by Saar \& Baliunas 1992): how directly and how far does the
solar analogy apply to other stars, and how do the underlying physical
processes differ?  Unfortunately, while stellar coronal X-ray emission
has been known and studied for more than 20 years, the small number
of satellites in orbit at any given time able to observe it severely
limits our knowledge of any long-term trends in stellar X-ray activity.
Such knowledge is currently restricted to a handful of stars caught during
repeated brief snapshots of them afforded by observations of different
satellites.

If magnetic cycles with similar timescales to that of the Sun are
present on other stars, as convincing evidence from the long-term
Mt.~Wilson Ca\,II H+K monitoring program suggests (e.g., Baliunas et
al.\ 1995 and references therein), then one might also expect these
stars to modulate their coronal X-ray fluxes in a similar way to the
Sun.  Further, on the Sun these modulations are large: Solrad 
observations (Kreplin
1970) in the 44-60\AA\  and 8-20\AA\ passbands show that X-ray
flux at activity maximum (c.1968) is $\sim 20$ and $\gtrsim 200$ times
greater than at activity minimum (July 1964) respectively (see also
Vaiana \& Rosner 1978).  Also, as stated by Hempelmann, Schmitt \&
St\c{e}pi\`en (1996), the variation of the solar X-ray flux in the
equivalent of the {\it ROSAT}/PSPC bandpass over its activity cycle
is a factor of 10 or more (also Pallavicini 1993; but Ayres et al.\ 1996,
extrapolating from XUV data predict a variation by only a factor $\sim
4$), similar to the ratio deduced for the variations in the soft X-ray
range of {\it Yohkoh} based on ratios of X-ray fluxes in the 1-8\AA\ 
passband of GOES (Aschwanden 1994), and as also directly observed by
{\it Yohkoh} (Acton 1996).  Such large long-term changes in
mean stellar X-ray flux levels are, at least in principle, easily detectable.
However, studies of stellar X-ray emission at different epochs based on
{\it Einstein} and subsequent {\it ROSAT} observations of stars in open
clusters (Stern et al.\ 1995, Gagn\'{e} et al.\ 1995; Micela et al.\
1996), as well as field stars (Schmitt, Fleming, \& Giampapa
1995, Fleming et al.\ 1995) suggest that these active stars, at least, do
not show strong long-term components of variability; some of these
results are discussed by Stern (1998).
A recent study by Hempelmann et al.\ (1996) of F-K main sequence stars
also finds that the more active stars with higher surface X-ray fluxes
tend not to have well-defined cyclic activity in terms of the Ca II H+K
activity index.  Some authors have suggested that this lack of clear
detection of activity cycles might be an observational consequence of the
dominant magnetic activity on the more active stars being due to a different
dynamo process to the solar large-scale field $\alpha\,\omega$ dynamo
(e.g., Stern et al.\ 1995; Drake et al.\ 1996).

In this paper, we turn to the most active stars -- the RS\,CVn and 
BY\,Dra binaries -- in order to investigate whether or not they might
exhibit some form of cyclic, or other long-term variability in their
X-ray emission.  We look at a sample of active binary stars that have
been detected by the {\it Einstein Observatory} (c.1978-81) and that
have also been observed by the {\it ROSAT}/PSPC both during the all-sky
survey (c.1990), and during later pointed observations (c.1991-1994).  We
compare the different observations in order to assess whether or not
there is any significant difference between {\em changes} in flux
levels over short-term timescales ($\sim \frac{1}{2}-2$ yrs; {\it ROSAT}
All-Sky Survey v/s pointed phase) compared with changes over longer-term
timescales ($\sim 10-12$ yrs; {\it ROSAT} v/s {\it Einstein}).

In \S\ref{s:sample} we describe the star sample used in this study.
In \S\ref{s:analyz} we describe the statistical method we adopt to
compare the samples and discuss the implications of our results: in
\S\ref{s:correl} we consider the correlations of the samples and their
deviations from equality; in \S\ref{s:signify} we discuss the statistical
significance of the analysis; and in \S\ref{s:cyclic} we discuss the
implications of our results in the context of stellar activity cycles.
We summarize in \S\ref{s:summary}.

\section{\label{s:sample}Data Selection}

We adopt the sample of 206 spectroscopic binary systems of Strassmeier et
al.\ (1993) as our baseline database of active stars.  This sample was
selected by Strassmeier et al.\ such that each system has at least one
late-type component that shows Ca\,II H and K emission in its spectrum.

In Table~\ref{t:starlist} we list a subset of the Strassmeier et
al.\ stars which have at least one X-ray measurement with either the
{\it Einstein}/IPC or the {\it ROSAT}/PSPC, together with the relevant
observed count rates.  We have used the widely available catalogs of
the {\it Einstein} Slew Survey (``Slew''; Elvis et al.\ 1992, Plummer
et al.\ 1994), the {\it Einstein} Extended Medium Sensitivity Survey
(``EMSS''; Gioia et al.\ 1990, Stocke et al.\ 1991), and the {\it
Einstein}/IPC Source Catalog (``EOSCAT''; Harris et al.\ 1990) to
obtain {\it Einstein}/IPC measurements (``Einstein''); and the {\it
ROSAT} All-Sky Survey (RASS) Bright Source Catalog (``RASSBSC''; Voges
et al.\ 1996) and {\it ROSAT} public archive pointed data sets
(``WGACAT''; White, Giommi, \& Angelini 1994) to obtain the PSPC
measurements.  We have not augmented the WGACAT with independently
measured fluxes in order to keep the X-ray sample
homogeneous.\footnote{Using
other existing catalogs (e.g., ``ROSATSRC''; Voges et al.\ 1994) of
{\it ROSAT} pointed data sets does not change the overlaps (cf.\
Table~\ref{t:tests}) significantly.
}

If a particular star is found in more than one {\it Einstein} survey
catalog, we adopt the count rate derived in EOSCAT over that of EMSS
over that of Slew.  If multiple PSPC pointings exist of a star, then
we use only the measurement with the highest effective exposure
(including vignetting) and the one closest to the field-center.

In comparing {\it Einstein}/IPC counts with {\it ROSAT}/PSPC counts
of the same star, we adopt a conversion factor $\frac{PSPC}{IPC} = 3.7$
based on a straight line fit to the Einstein-RASSBSC sample.  Clearly,
this is an approximate number that could vary according to the adopted
plasma temperature, the metallicity of the corona, and column density
of absorption to the source.  The bandpasses and effective areas of
both instruments are however similar enough over the temperature range
of interest that the ratio of count rates are insensitive to these
parameters (see \S\ref{s:signify}).  RASSBSC and WGACAT counts are
extracted in slightly different passbands, and an appropriate correction
($\sim 20\%$) has also been applied to these datasets.

{\tiny
{\tiny

\begin{table}[tb]
\tablenum{1}
\begin{center}
\caption{\label{t:starlist} Active Binaries observed in X-rays}
\begin{tabular}{r l c c c c c c r}
\hfil & \hfil & \hfil & \hfil & \hfil & \hfil & \hfil & \hfil & \hfil \\
\hline
Number$^1$ & Name & \multicolumn{3}{c}{\it Einstein} & \multicolumn{2}{c}{\it
ROSAT} & Spectral & Distance$^3$ \\
\hfil & \hfil & Slew$^2$ & EMSS$^2$ & EOSCAT$^2$ & RASSBSC$^2$ & WGACAT$^2$
& Type$^1$ & [pc] \\
\hline

  3 & 5 Cet  & ...  & ...  &     5.6 & ...  & ...  & wF/K1III & 307.7 \\
  4 & BD Cet  & ...  & ...  & ...  &    65.5 & ...  & K1III & 416.7 \\
  5 & 13 Cet A  & ...  & ...  & ...  &  1116.0 & ...  & {F7V/}G4V & 21.0 \\
  6 & FF And  & ...  & ...  & ...  &  1022.0 & ...  & dM1e/dM1e & 23.8 \\
  7 & $\zeta$ And  &   650.0 & ...  &   467.0 &  1797.0 & ...  & /K1III & 55.6 \\
  8 & CF Tuc  &   300.0 & ...  &   394.0 & ...  & ...  & G0V/K4IV & 86.2 \\
  9 & BD+25 161  & ...  & ...  & ...  &   240.9 & ...  & G2V & 215.5 \\
 10 & AY Cet  &  1850.0 & ...  &  1516.0 &  3939.0 & ...  & WD/G5III & 78.5 \\
 11 & UV Psc  &   250.0 & ...  &   219.0 &   924.5 & ...  & G4-6V/K0-2V & 63.0 \\
 12 & CP-57 296  & ...  & ...  & ...  &   491.6 & ...  & G6-8IV-IIIe & 117.8 \\
 13 & BI Cet  &   340.0 & ...  & ...  &  1520.0 & ...  & G5V:/G5V: & 65.7 \\
 14 & AR Psc  & ...  & ...  &   686.0 &  4673.0 & ...  & K2V/? & 45.2 \\
 15 & UV For  & ...  & ...  & ...  &   140.9 & ...  & K0IV & 130.4 \\
 16 & BD+34 363  & ...  & ...  & ...  &   995.5 & ...  & K0III & 196.9 \\
 17 & 6 Tri  &   460.0 & ...  &   416.0 &  1023.0 & ...  & F5/K0III & 93.6 \\
 20 & UX For  &  2220.0 & ...  & ...  &  1841.0 & ...  & G5-8V/(G) & 40.4 \\
 22 & VY Ari  &   960.0 & ...  &   772.0 &  6792.0 &  3790.0 & K3-4V-IV & 44.0 \\
 23 & BD+25 497  & ...  & ...  & ...  &    52.3 & ...  & G4V/G6V & 77.5 \\
 24 & EL Eri  & ...  & ...  & ...  &   443.0 & ...  & G8IV-III & 219.3 \\
 25 & LX Per  & ...  & ...  &   129.0 &   571.4 & ...  & G0IV/K0IV & 100.0 \\
 28 & UX Ari  &  3500.0 & ...  &  4360.0 &  5859.0 &  7710.0 & G5V/K0IV & 50.2 \\
 29 & V711 Tau  &  4020.0 & ...  &  4142.0 & 18970.0 &  8950.0 & G5IV/K1IV & 29.0 \\
 30 & V837 Tau  & ...  & ...  & ...  &  3996.0 & ...  & G2V/K5V & 37.3 \\
 31 & HR 1176  & ...  & ...  & ...  &    83.9 & ...  & F2:V/G8III & 106.3 \\
 33 & CF Tau  & ...  & ...  &    21.5 & ...  & ...  & F8 & 200$^1$ \\
 34 & AG Dor  & ...  & ...  & ...  &   839.4 & ...  & K1Vp & 34.9 \\
 35 & EI Eri  &  1980.0 & ...  &  1394.0 &  4409.0 & ...  & G5IV & 56.2 \\
 37 & V818 Tau  & ...  & ...  &   520.0 &   236.9 &   218.0 & G6V/K6V & 46.7 \\
 38 & BD+17 703  & ...  & ...  & ...  &   109.3 &    57.9 & G4V/G8V & 45.8 \\
 39 & BD+14 690  &    90.0 & ...  &    94.1 &   197.1 &   123.0 & G0V & 46.6 \\
 40 & vB 69  & ...  & ...  & ...  & ...  &    43.1 & K0V & 49.8 \\
 41 & V492 Per  & ...  & ...  & ...  &   230.6 & ...  & K1III & 118.1 \\
 42 & V833 Tau  &   730.0 & ...  &   524.0 &  2689.0 &  1380.0 & dK5e & 17.9 \\
 43 & 3 Cam  & ...  & ...  &    53.1 &    94.7 & ...  & K0III & 152.0 \\
 44 & RZ Eri  &   120.0 & ...  &   109.4 &   238.2 &   265.0 & Am/K0IV & 185.2 \\
 45 & V808 Tau  & ...  & ...  & ...  &   191.5 &   159.0 & K3V/K3V & 52.7 \\
 47 & V1198 Ori  & ...  & ...  & ...  &  1364.0 & ...  & G5IV & 33.5 \\
 48 & 12 Cam  &   350.0 & ...  &   207.0 &   466.8 &   673.0 & K0III & 191.6 \\
 49 & HP Aur  & ...  & ...  & ...  &    52.5 & ...  & G8 &  ...  \\
 50 & CP-77 196  & ...  & ...  & ...  &   376.3 & ...  & K1IIIp & 179.5 \\
 51 & $\alpha$ Aur  & ...  & ...  &  4222.0 & 23540.0 & 13300.0 & G1III/K0III & 12.9 \\
 52 & BD+75 217  & ...  & ...  & ...  &   205.1 & ...  & K0III & 136.8 \\
 54 & TW Lep  & ...  & ...  & ...  &  1609.0 & ...  & F6IV/K2III & 170.4 \\
 55 & V1149 Ori  &   200.0 & ...  &   148.5 &   626.1 & ...  & K1III & 144.3 \\
 56 & CD-28 2525  & ...  & ...  & ...  &   166.7 & ...  & G1V & 87.6 \\
 57 & HR 2072  & ...  & ...  & ...  &   654.8 & ...  & F/G5-8III & 127.2 \\
 58 & SZ Pic  & ...  & ...  & ...  &   454.2 & ...  & G8V & 194.9 \\
 59 & HR 2054  & ...  & ...  & ...  &   607.0 &   417.0 & G8III & 178.9 \\
 60 & CQ Aur  & ...  & ...  &    50.8 & ...  & ...  & F5/K1IV & 242.1 \\
 61 & TY Pic  & ...  & ...  & ...  &   219.0 & ...  & F/G8-K0III & 286.5 \\

\hline
\end{tabular}
\end{center}
\end{table}

\begin{table}[htb]
\tablenum{1}
\begin{center}
\caption{(Contd.) Active Binaries observed in X-rays}
\begin{tabular}{r l c c c c c c r}
\hfil & \hfil & \hfil & \hfil & \hfil & \hfil & \hfil & \hfil & \hfil \\
\hline
Number$^1$ & Name & \multicolumn{3}{c}{\it Einstein} & \multicolumn{2}{c}{\it
ROSAT} & Spectral & Distance$^3$ \\
\hfil & \hfil & Slew$^2$ & EMSS$^2$ & EOSCAT$^2$ & RASSBSC$^2$ & WGACAT$^2$
& Type$^1$ & [pc] \\
\hline

 62 & OU Gem  &   230.0 & ...  &   287.0 &  1741.0 & ...  & K3V/K5V & 14.7 \\
 63 & TZ Pic  & ...  & ...  & ...  &   165.5 & ...  & K1IV-IIIp & 175.7 \\
 64 & W92/NGC2264  & ...  & ...  &    22.7 &   137.7 &    55.9 & K0:IVp & 900$^1$ \\
 65 & SV Cam  & ...  & ...  & ...  &   408.0 &   509.0 & G2-3V/K4V & 85.0 \\
 66 & VV Mon  & ...  & ...  &    34.4 &   147.2 & ...  & G2IV/K0IV & 178.9 \\
 67 & Gl 268  & ...  & ...  &    40.8 &   257.0 & ...  & dM5e/dM5e & 6.4 \\
 68 & SS Cam  & ...  & ...  &    37.0 &    58.5 & ...  & F5V-IV/K0IV-III & 323.6 \\
 69 & HR 2814  & ...  & ...  & ...  &   353.7 & ...  & F9.5V{K3:V/(K5V} & 34.8 \\
 70 & AR Mon  & ...  & ...  &    42.1 &    97.1 & ...  & G8III/K2-3III & 276.2 \\
 71 & YY Gem  &   900.0 & ...  &   464.0 &  3697.0 &  2570.0 & dM1e/dM1e & 15.8 \\
 72 & V344 Pup  & ...  & ...  & ...  &   214.8 & ...  & K1III & 111.4 \\
 73 & $\sigma$ Gem  &  2090.0 & ...  &  1557.0 &  8081.0 &  7260.0 & K1III & 37.5 \\
 75 & 54 Cam  &   250.0 & ...  &   227.7 &  1347.0 & ...  & F9IV/G5IV & 101.6 \\
 76 & LU Hya  & ...  & ...  & ...  &   184.1 & ...  & K1IV & 49.5 \\
 78 & HR 3385  & ...  & ...  & ...  &   714.6 & ...  & K0III & 130.7 \\
 79 & RU Cnc  & ...  & ...  &    23.1 &   107.3 & ...  & F5IV/K1IV & 331.1 \\
 80 & RZ Cnc  &   140.0 & ...  &   149.2 &    66.1 & ...  & K1III/K3-4III & 307.7 \\
 81 & TY Pyx  & ...  & ...  & ...  &  1604.0 &  1930.0 & G5IV/G5IV & 55.8 \\
 83 & XY UMa  & ...  & ...  & ...  & ...  &     2.9 & G3V/(K4-5V) & 150.8 \\
 84 & BF Lyn  &   860.0 & ...  & ...  &  2908.0 & ...  & K2V/(dK) & 24.3 \\
 85 & IL Hya  & ...  & ...  & ...  &  1703.0 & ...  & K1III & 119.6 \\
 86 & IN Vel  & ...  & ...  & ...  &   120.8 & ...  & K2IIIp & 285.7 \\
 87 & DH Leo  &   970.0 & ...  & ...  &  2053.0 & ...  & {K0V/K7V}K5V & 32.4 \\
 88 & XY Leo B  & ...  & ...  &    96.9 &   368.1 &   258.0 & M1V/M3V & 63.1 \\
 89 & BD+61 1183  & ...  & ...  & ...  &   347.8 & ...  & G8IV & 174.5 \\
 90 & LR Hya  & ...  & ...  & ...  &    79.9 &    64.0 & K0V/K0V & 33.8 \\
 91 & DM UMa  &   570.0 & ...  & ...  &   897.1 & ...  & K0-1IV-III & 138.7 \\
 92 & $\xi$ UMa B  &  2580.0 & ...  &   873.0 &  4539.0 & ...  & G5V & 8$^1$ \\
 95 & CD-38 7259  & ...  & ...  & ...  &   928.6 & ...  & G5V/K1IV & 121.7 \\
 96 & HR 4492  &  1290.0 & ...  & ...  &  2959.0 & ...  & A0/K2-4III & 172.1 \\
 97 & RW UMa  &   170.0 & ...  &    89.3 &    78.2 & ...  & F8IV/K0IV & 242.1 \\
 98 & 93 Leo  &   410.0 & ...  &   259.0 &  1122.0 & ...  & A6:V/G5IV-III & 69.4 \\
 99 & HU Vir  & ...  & ...  & ...  &   428.9 & ...  & K0IV & 125.0 \\
100 & HR 4665  &   580.0 & ...  &   606.0 &  1638.0 & ...  & K1III/K1III & 138.1 \\
101 & AS Dra  & ...  & ...  &    80.0 &   236.4 & ...  & G4V/G9V & 43.2 \\
102 & IL Com  & ...  &    68.1 &    71.0 &   600.1 &   488.0 & F8V/F8V & 107.1 \\
103 & BD+25 2511  & ...  & ...  & ...  &   201.2 &   241.0 & <G9V> & 55$^1$ \\
104 & BD-05 3578  & ...  &    63.6 &    28.0 &   228.8 &   203.0 & G5V/(K-M) & 1204.8 \\
105 & IN Com  & ...  & ...  & ...  & ...  &     6.2 & G5IV-III & 190.8 \\
106 & BD+47 2007  & ...  & ...  & ...  &   395.0 &   381.0 & F/K0III & 190.8 \\
107 & UX Com  &   120.0 & ...  &    50.2 &   196.5 & ...  & G2/K1(IV) & 168.4 \\
108 & BD-4 3419  & ...  & ...  & ...  &   193.6 & ...  & K2IV-III & 300.3 \\
109 & RS CVn  &   380.0 & ...  &   390.0 &   886.2 &   634.0 & F4IV/G9IV & 108.1 \\
110 & HR 4980  & ...  & ...  & ...  &   756.7 & ...  & G0V/G0V & 39.8 \\
111 & BL CVn  & ...  & ...  & ...  &    71.5 & ...  & G-KIV/K0III & 284.9 \\
112 & BM CVn  & ...  & ...  & ...  &  1316.0 & ...  & K1III & 111.1 \\
114 & HR 5110  &  1200.0 & ...  &  1542.0 &  2768.0 &  4580.0 & F2IV/K2IV & 44.5 \\
115 & CD-32 9477  & ...  & ...  &    46.3 &   127.0 &   144.0 & K2IIIp & 507.6 \\
116 & V851 Cen  &   680.0 & ...  &   551.0 &  1303.0 & ...  & K2IV-III & 76.2 \\
117 & BH Vir  & ...  & ...  & ...  &   164.9 & ...  & F8V-IV/G2V & 125.9 \\

\hline
\end{tabular}
\end{center}
\end{table}

\begin{table}[htb]
\tablenum{1}
\begin{center}
\caption{(Contd.) Active Binaries observed in X-rays}
\begin{tabular}{r l c c c c c c r}
\hfil & \hfil & \hfil & \hfil & \hfil & \hfil & \hfil & \hfil & \hfil \\
\hline
Number$^1$ & Name & \multicolumn{3}{c}{\it Einstein} & \multicolumn{2}{c}{\it
ROSAT} & Spectral & Distance$^3$ \\
\hfil & \hfil & Slew$^2$ & EMSS$^2$ & EOSCAT$^2$ & RASSBSC$^2$ & WGACAT$^2$
& Type$^1$ & [pc] \\
\hline

118 & V841 Cen  & ...  & ...  & ...  &   800.8 &   669.0 & K1IV & 63$^1$ \\
119 & RV Lib  & ...  & ...  &    60.4 &   160.5 & ...  & G8IV/K3IV & 370.4 \\
120 & HR 5553  & ...  & ...  &    78.0 &   362.9 &   294.0 & K2V & 11.5 \\
121 & SS Boo  & ...  & ...  &    47.3 &    64.0 &    75.3 & G0V/K0IV & 202.0 \\
122 & UV CrB  & ...  &    18.1 &    17.9 & ...  & ...  & K2III & 279.3 \\
123 & GX Lib  & ...  &    34.1 &    33.3 &   138.1 & ...  & (G-KV)/K1III & 95.1 \\
124 & LS TrA  & ...  & ...  & ...  &  1189.0 & ...  & K2IV/K2IV & 127.4 \\
126 & RT CrB  & ...  & ...  &    10.7 & ...  & ...  & G2/G5-8IV & 1428.6 \\
128 & 1E1548.7+1125  & ...  &    28.8 &    28.4 & ...  &   127.0 & K5V-IV & 500$^1$ \\
130 & MS Ser  &   300.0 & ...  &    85.1 &   642.6 & ...  & K2V/K6V & 87.8 \\
131 & BD+11 2910  & ...  & ...  & ...  &   359.2 & ...  & G8IV & 40.3 \\
132 & $\sigma^2$ CrB  &  2270.0 & ...  &  2170.0 &  9487.0 &  8710.0 & F6V/G0V & 21.7 \\
134 & CM Dra  & ...  & ...  & ...  &   176.7 & ...  & M4Ve/M4Ve & 15$^1$ \\
136 & WW Dra  &   350.0 & ...  &    84.2 &   493.3 & ...  & G2IV/K0IV & 115.3 \\
137 & $\epsilon$ UMi  & ...  & ...  & ...  &  1048.0 & ...  & A8-F0V/G5III & 106.3 \\
138 & CD-26 11634  & ...  & ...  & ...  &   372.3 & ...  & K0III & 471.7 \\
139 & V792 Her  & ...  & ...  &   184.0 &   365.6 & ...  & F2IV/K0III & 413.2 \\
141 & V824 Ara  & ...  & ...  & ...  &  4758.0 & ...  & G5IV/K0V-IV & 31.4 \\
142 & HR 6469  & ...  & ...  & ...  &   174.9 & ...  & {F2V/ G0V }G5IV & 64.4 \\
143 & V965 Sco  & ...  & ...  & ...  &   200.3 &   144.0 & F2IV/K1III & 406.5 \\
144 & 29 Dra  &   800.0 & ...  & ...  &  1272.0 &  2320.0 & WD/K0-2III & 103.3 \\
146 & BD+36 2975  & ...  & ...  & ...  &  1875.0 & ...  & G6V/K1IV & 30.9 \\
147 & Z Her  & ...  & ...  &   113.0 & ...  & ...  & F4V-IV/K0IV & 98.3 \\
148 & MM Her  & ...  & ...  &    64.3 &   133.7 & ...  & G2/K0IV & 184.5 \\
149 & V772 Her  & ...  & ...  &   542.9 &  2886.0 &  1890.0 & {G0V/ M1V }G5V & 37.7 \\
150 & ADS 11060C  & ...  & ...  &   542.9 &  2886.0 &  1890.0 & K7:V/K7V & 37.7 \\
151 & V832 Ara  & ...  & ...  & ...  &   233.9 &   237.0 & WD/G8III & 266.7 \\
152 & V815 Her  & ...  & ...  & ...  &  3113.0 & ...  & G5V/(M1-2V) & 32.6 \\
153 & PW Her  & ...  & ...  &    43.2 &   116.7 & ...  & F8-G2/K0IV & 232.0 \\
155 & AW Her  & ...  & ...  &    35.6 &    88.4 & ...  & G2/G8IV & 212.3 \\
156 & BY Dra  &  1020.0 & ...  &   556.0 &  2414.0 &  1460.0 & K4V/K7.5V & 16.4 \\
157 & 1E1848+3305  & ...  & ...  &    51.6 &   166.7 & ...  & K0III-IV & 229$^1$ \\
158 & 1E1848+3325  & ...  & ...  &    16.7 & ...  & ...  & G5V & 95$^1$ \\
159 & $o$ Dra  & ...  & ...  &    16.6 &    50.5 & ...  & G9III & 98.8 \\
160 & V1285 Aql  & ...  & ...  &   159.5 &  1392.0 & ...  & M3.5Ve/M3.5Ve & 11.6 \\
161 & V775 Her  & ...  & ...  & ...  &  1878.0 &  1640.0 & K0V/(K5-M2V) & 21.4 \\
162 & V478 Lyr  & ...  & ...  & ...  &  2095.0 & ...  & G8V/(dK-dM) & 28.0 \\
163 & HR 7275  &   810.0 & ...  &   788.0 &  1275.0 & ...  & K1IV-III & 70.2 \\
164 & 1E1919+0427  & ...  & ...  &    29.1 &   126.5 & ...  & G5V/K0III-IV & 95$^1$ \\
165 & V4138 Sgr  & ...  & ...  & ...  &  1345.0 & ...  & K1III & 87.7 \\
166 & V4139 Sgr  & ...  & ...  & ...  &   100.3 & ...  & K2-3III & 240.4 \\
167 & HR 7428  & ...  & ...  &    76.0 &   107.9 &   131.0 & A2V/K2III-II & 322.6 \\
169 & 1E1937+3027  & ...  & ...  &    21.5 & ...  &    88.8 & K0III-IV & 229$^1$ \\
171 & HR 7578  & ...  & ...  & ...  &   415.9 & ...  & K2-3V/K2-3V & 14.2 \\
173 & BD+15 4057  & ...  & ...  & ...  &   179.7 & ...  & G5V/G5V & 55.8 \\
174 & BD+31 4046  & ...  & ...  & ...  &   239.0 & ...  & K0III & 275$^1$ \\
175 & BI Del  & ...  & ...  & ...  &    82.7 & ...  & K0 &  ...  \\
176 & AT Cap  & ...  & ...  & ...  &    67.1 & ...  & K2III & 99$^1$ \\
177 & CG Cyg  & ...  & ...  & ...  &   143.1 & ...  & G9.5V/K3V & 108.1 \\
178 & V1396 Cyg  &   410.0 & ...  & ...  &  1013.0 & ...  & M2V/M4Ve & 15.1 \\

\hline
\end{tabular}
\end{center}
\end{table}

\begin{table}[htb]
\tablenum{1}
\begin{center}
\caption{(Contd.) Active Binaries observed in X-rays}
\begin{tabular}{r l c c c c c c r}
\hfil & \hfil & \hfil & \hfil & \hfil & \hfil & \hfil & \hfil & \hfil \\
\hline
Number$^1$ & Name & \multicolumn{3}{c}{\it Einstein} & \multicolumn{2}{c}{\it
ROSAT} & Spectral & Distance$^3$ \\
\hfil & \hfil & Slew$^2$ & EMSS$^2$ & EOSCAT$^2$ & RASSBSC$^2$ & WGACAT$^2$
& Type$^1$ & [pc] \\
\hline

179 & ER Vul  &   570.0 & ...  &   391.0 &  2310.0 &  1670.0 & G0V/G5V & 49.9 \\
181 & BD+10 4514  & ...  & ...  & ...  &   262.7 & ...  & {F9V/G0V}GIV & 50.5 \\
182 & HR 8170  & ...  & ...  & ...  &   729.2 &   638.0 & F8V/wK5V & 26.6 \\
183 & BH Ind  & ...  & ...  & ...  &   345.6 & ...  & K1IIICNIVp & 310.6 \\
184 & BD-00 4234  & ...  & ...  & ...  &   388.6 & ...  & K3Ve/K7Ve & 49.4 \\
185 & AS Cap  & ...  & ...  & ...  &   326.5 & ...  & K1III & 204.1 \\
186 & AD Cap  & ...  & ...  &   137.0 &   305.6 & ...  & G5-8IV-V/G5 & 191.6 \\
187 & 42 Cap  & ...  & ...  & ...  &   639.0 & ...  & G2IV & 32.5 \\
188 & FF Aqr  & ...  & ...  & ...  &  5346.0 &   335.0 & sdO-B/G8IV-III & 126.4 \\
189 & RT Lac  &   120.0 & ...  &   110.5 &   141.9 &   224.0 & G5:/G9IV & 192.7 \\
190 & HK Lac  &   300.0 & ...  &   357.0 &  1463.0 & ...  & F1V/K0III & 151.1 \\
191 & AR Lac  &  1860.0 & ...  &  1268.0 &  7786.0 &  4010.0 & G2IV/K0IV & 42.0 \\
192 & WW Cep  & ...  & ...  & ...  &   158.3 & ...  & K0 &  ...  \\
193 & BD+29 4645  & ...  & ...  & ...  &   932.6 & ...  & F5-8/G8IV & 145.1 \\
194 & V350 Lac  & ...  & ...  &   163.0 &   670.0 & ...  & K2III & 122.2 \\
195 & FK Aqr  &   660.0 & ...  &   787.0 &  3803.0 & ...  & dM2e/dM3e & 8.6 \\
196 & IM Peg  & ...  & ...  &   585.0 &  2505.0 &  3730.0 & K2III-II & 96.8 \\
198 & TZ PsA  &   450.0 & ...  & ...  &   792.4 & ...  & G5Vp & 65.8 \\
199 & KU Peg  & ...  & ...  & ...  &   173.2 & ...  & G8II & 187.6 \\
200 & KZ And  &   540.0 & ...  &   366.0 &  1928.0 &  1540.0 & dK2/dK2 & 25.3 \\
201 & RT And  & ...  & ...  & ...  &   224.2 & ...  & F8V/K0V & 75.4 \\
202 & SZ Psc  & ...  & ...  &   976.0 &  4463.0 & ...  & F8IV/K1IV & 88.2 \\
203 & EZ Peg  & ...  & ...  & ...  &   679.1 &   438.0 & G5V-IV/K0IV: & 129.5 \\
204 & $\lambda$ And  &  2850.0 & ...  &  2080.0 &  9822.0 &  6000.0 & G8IV-III & 25.8 \\
205 & KT Peg  & ...  & ...  & ...  &   363.7 & ...  & G5V/K6V & 49.3 \\
206 & II Peg  &  2980.0 & ...  &  1156.0 & 10900.0 &  9040.0 & K2-3V-IV & 42.3 \\

\hline
\multicolumn{9}{l}{NOTES:-} \\
\multicolumn{9}{l}{1. as in Strassmeier et al.\ (1993).} \\
\multicolumn{9}{l}{2. Vignetting-corrected count-rates [ct ks$^{-1}$].} \\
\multicolumn{9}{l}{3. from Hipparcos catalog (Perryman et al.\ 1997)} \\
\end{tabular}
\end{center}
\end{table}

}

}

\section{\label{s:analyz}Analysis and Discussion}

Subsets of the active binary stars that have been observed in two different
epochs allow us to deduce the magnitude of the variability at different
timescales.  We begin by assuming that each such sample is statistically
random, i.e., that there are no systematic changes in the variability of
the sample from one epoch to the other, or in other words, that on average
any increases in intrinsic luminosity is balanced by decreases in intrinsic
luminosity.  This assumption is supported by the Kruskal-Wallis test for
both the combined samples (i.e., active binaries with X-ray data at all 3
epochs) and for the individual samples (active binaries observed in any of
the 3 epochs): the hypothesis that the samples have the same mean cannot
be rejected (the probability of obtaining the observed value of the K-W
test statistic by chance is $0.84 \pm 0.02$ and $0.4 \pm 0.06$ respectively,
much higher than an acceptable threshold of 0.05).\footnote{Fleming et
al.\ (1995) noted an increase in the mean X-ray emission level of a small,
X-ray selected, sample of RS\,CVn and W\,UMa binaries between the {\it
Einstein} and {\it ROSAT} epochs.  However, the sample of stars used here
has very little overlap with the Fleming et al.\ sample and is furthermore
much larger in size and significant changes in the observed mean X-ray
emission level are not expected.}
This result also confirms that the conversion factors correcting the
passband differences between the catalogs have been properly evaluated.

We ignore censored data (stars detected in one survey but not in another
[11 in Einstein, 3 in WGACAT, 65 in RASSBSC],
as well as stars not detected in any survey [29]) in this work.  The large
dynamic range of the observed count rates ($> 10^2$; cf.\
Figures~\ref{f:einrass}-\ref{f:rasswga}) in the samples, and the strong
correlations in the {\em detected} count rates show that ignoring
undetected stars will not affect the results presented here.  Further
note that we use a sample that is {\em not} X-ray selected, and thus
avoid the problem encountered by Fleming et al.\ (1995) who found
general decreases in overall flux between {\it Einstein} and {\it ROSAT}
measurements due to preferential detection of stars while flaring.

In the following sections we analyze the paired samples in greater
detail (\S\ref{s:correl}), assess the significance of our results
(\S\ref{s:signify}), and discuss the results in the context of stellar
activity cycles (\S\ref{s:cyclic}).

\subsection{\label{s:correl}Correlations and Deviations}

\begin{figure*}
\plotone{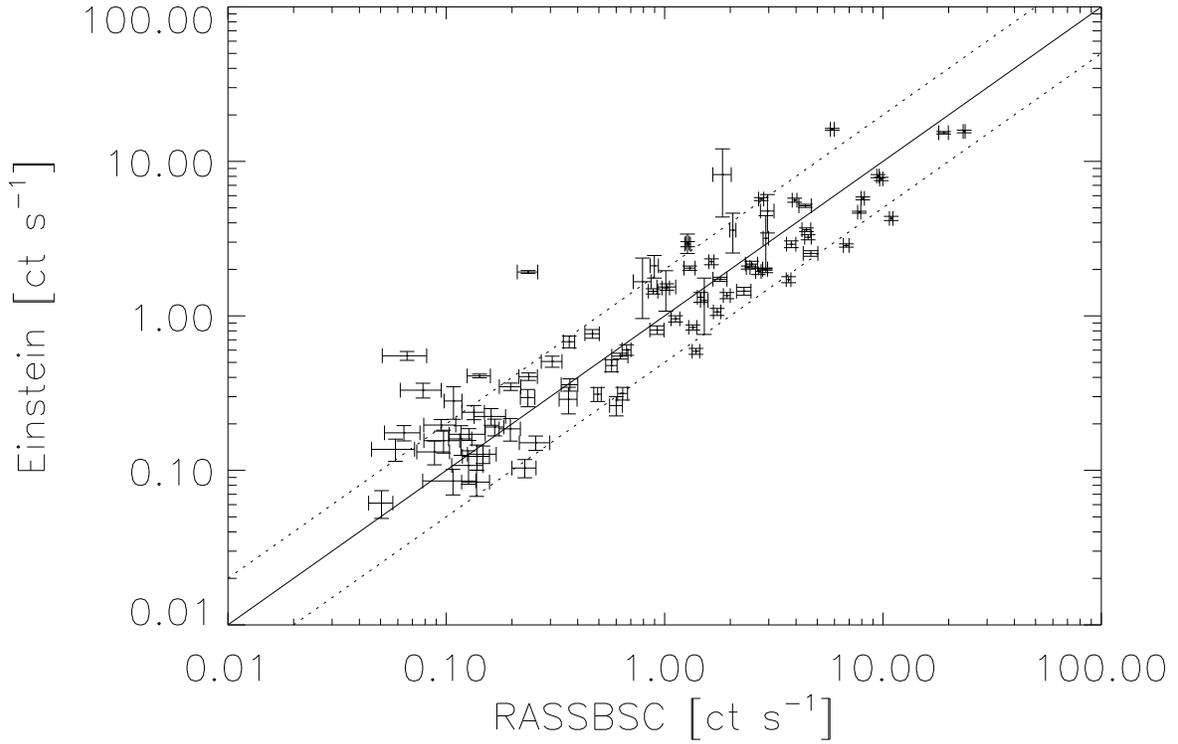}
\caption{Scatter plot of count rates observed at different epochs:
RASSBSC v/s Einstein.
The Einstein count rates have been uniformly multiplied by a factor
of 3.7 as derived by a straight-line fit to the Einstein-RASSBSC dataset.
The size of the $1\sigma$ errors on the rates is indicated by the
horizontal and vertical lines at each point.  The solid inclined line
represents the line of equality and the dotted lines flanking it represent
variations of factors of 2.
\label{f:einrass}}
\end{figure*}

\begin{figure*}
\plotone{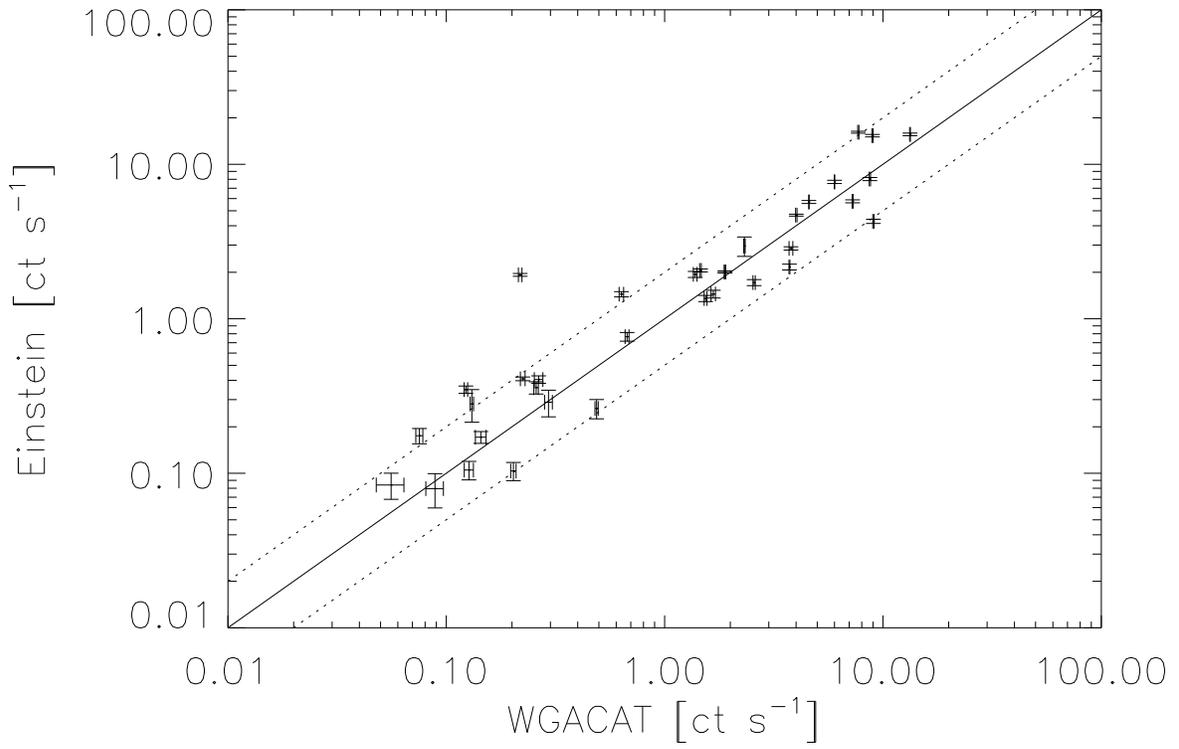}
\caption{As in Figure~\ref{f:einrass}, for the Einstein-WGACAT sample.
\label{f:einwga}}
\end{figure*}

\begin{figure*}
\plotone{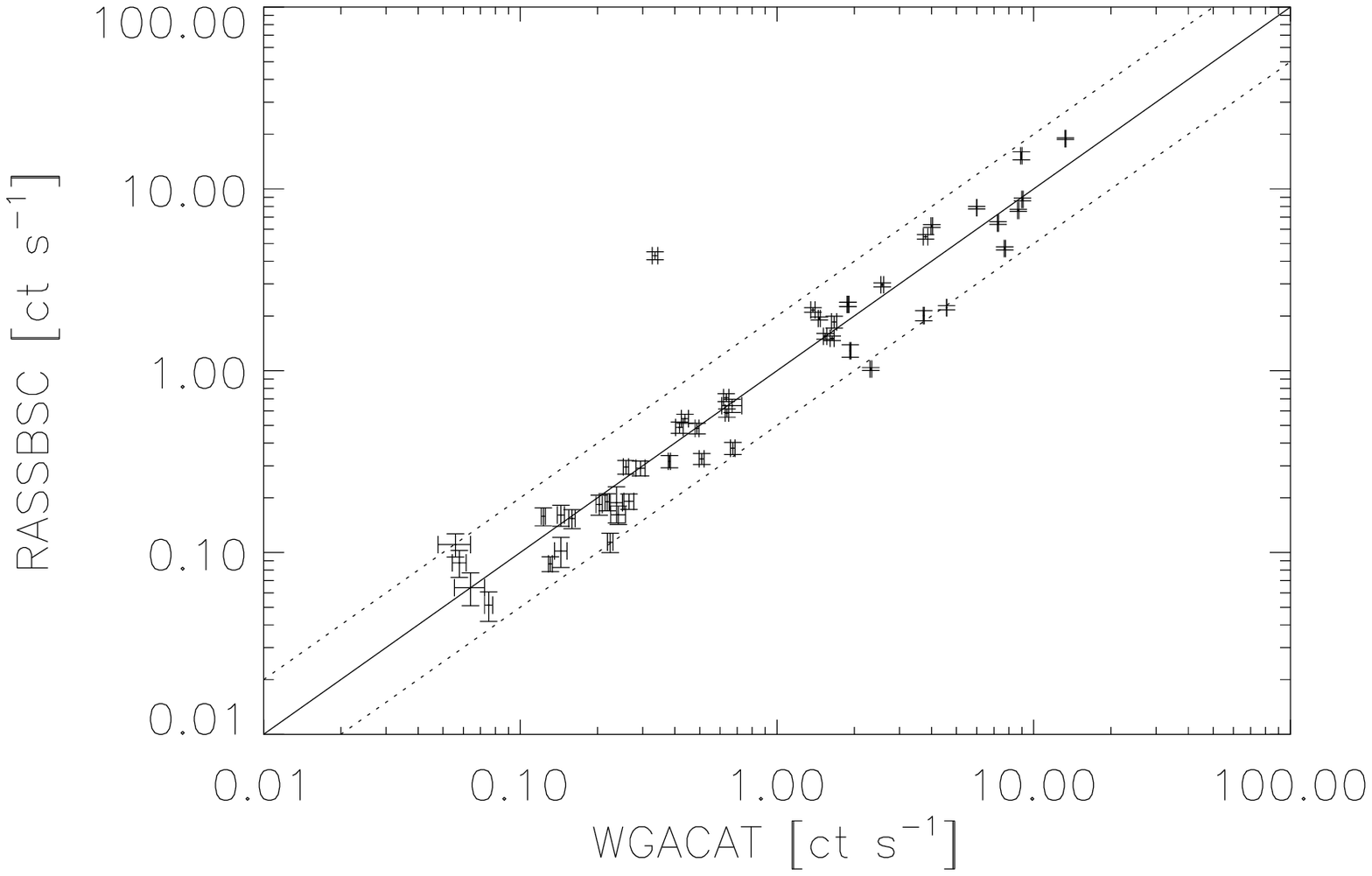}
\caption{As in Figure~\ref{f:einrass}, for the RASSBSC-WGACAT sample.
The scatter is much less pronounced except for the sole outlier, FF Aqr.
\label{f:rasswga}}
\end{figure*}

The sample of stars able to shed light on ``short'' timescale ($\sim
0-3$ yr) variability, i.e., stars re-observed after a short interval,
are those active binaries present in both the RASSBSC and WGACAT, while
there are two sets of paired datasets defining the ``long'' timescales
($\gtrsim 10$ yr) -- Einstein-RASSBSC and Einstein-WGACAT.  These
paired samples are shown in Figures~\ref{f:einrass}-\ref{f:rasswga}: it
is clear that the count rates are strongly correlated as one would
expect in the case where intrinsic variability of a single star is much
smaller than the range in brightness of the whole sample.  For
completeness, and to define the strength of the correlations in count
rates, we have performed standard statistical correlation tests, the
results of which are listed in Table~\ref{t:tests}.  We have tested the
sensitivity of the derived correlation coefficients to the statistical
errors on the observed count rates by performing monte carlo
simulations.  These involved generating a new set of count rates (of the
same sample size) for each star by sampling from a Gaussian with a mean
identical to the observed value and standard deviation equal to the
observed $1\sigma$ error; correlation coefficients are then derived using
the new set of simulated count rates.  We find that the derived coefficients
are stable to within $\lesssim 0.01$.

\begin{table}[ht]
\tablenum{2}
\begin{center}
\caption{\label{t:tests} Corelation Tests}
\begin{tabular}{l c c c}
\hfil & \hfil & \hfil & \hfil \\
\hline
\hfil & Einstein-RASSBSC & Einstein-WGACAT & RASSBSC-WGACAT \\
\hline
Sample Size		& 83		& 35 		& 47 \\
Pearson Coefficient	& 0.84		& 0.89 		& 0.94 \\
Spearman's $\rho$	& 0.91  	& 0.94  	& 0.97 \\
Kendall's $\tau$	& 0.73  	& 0.79  	& 0.86 \\
\hline
\end{tabular}
\end{center}
\end{table}

The strong correlations within the paired samples imply that any actual
variability in X-ray emission within the sample is not much larger than
the measurement errors.  Indeed, the majority of the observed
count rates in the different samples are within a factor of 2 of each
other (after allowing for the conversion between the {\it Einstein} and
{\it ROSAT} passbands).  This result is similar to other comparisons of {\it
Einstein} and {\it ROSAT} observations of samples of mostly active late-type
stars (e.g., Schmitt et al.\ 1995; Stern et al.\ 1995; Gagn\'{e} et al.\
1995; Micela et al.\ 1996; Fleming et al.\ 1995).
However, the larger scatter apparent in the Einstein-RASSBSC (and to a
lesser extent, Einstein-WGACAT) samples compared to the RASSBSC-WGACAT sample
(cf.\ Figures~\ref{f:einrass}-\ref{f:rasswga}) does appear to indicate
the presence of some non-statistical scatter in the data.  In the following,
we quantify this apparent variability.

The issue we seek to address is the extent of the departure of a
paired set of count rates from strict equality.  Further, any
measure of this departure must include the effects of the statistical
uncertainties associated with the observed count rates.  Thus, we
define the quantity
\begin{equation}
\label{e:dperp}
\delta_{\perp} = \frac{1}{N_{samp}}
	\sum_{samp} \frac{D_{\perp}}{\sigma_{tot}}
\end{equation}
where $D_{\perp}$ is the {\em perpendicular} distance of the pair of
count rates from a straight line of unit slope passing through the origin,
and $\sigma_{tot}$ is the total error associated with that pair as obtained
by propagating the individual errors, and $N_{samp}$ is the number of
paired count rates in the sample; if the count rates in the two samples
are identical $\delta_{\perp} = 0$, and in the case of only statistical
variations, $\delta_{\perp} \sim 1$.  Note that this is similar (but differs
in the use of perpendicular deviations and division by the error) to the
merit-function used to derive straight-line fits to data such that {\sl
absolute deviation} is minimized (cf.\ Press et al.\ 1992).  In the case
of small deviations, $D_{\perp}$ may be obtained from the logarithmic ratio
of the count rates.\footnote{This
is the statistic adopted by Gagn\'{e} et al. (1995).  Taking the count rates
observed at two epochs to be $c_1$ and $c_2$, with $c_1 = c_2 + \delta_{12}$,
${\rm ln}\left(\frac{c_1}{c_2}\right)$ $\sim$
${\rm ln}\left(1+\frac{\delta_{12}}{c_2}\right)$ $\sim$
$\frac{\delta_{12}}{c_2}$ $\sim$ $\frac{D_{\perp}}{c_2}$.
Note that this formulation preserves sign information (i.e., whether the
first or the second epoch has the higher count rate; the expectation value
of this statistic is 0 in the absence of variability, unlike that of
$\delta_{\perp}$ which has an expectation value $\sim 1$).  However,
since we are only interested in deviations from constancy, we essentially
marginalize over this two-sidedness (and thereby improve our detection
efficiency) by using the perpendicular deviates $D_{\perp}$.  Using the
perpendicular deviates also allows us to include the effects of the
measurement uncertainties in a straightforward fashion.
}
Note that standard statistical measures such as the Students T, the
F-statistic, the Sign test, etc.\ apply to the means and variances of the
samples, and are not sensitive enough for our purposes.  The adopted method
also has the advantage of allowing us to parameterize the detected
variability (albeit crudely; see \S\ref{s:signify}).
The values of $\delta_{\perp}$ derived from the three pairs of 
datasets as defined in Equation~\ref{e:dperp} are listed in
Table~\ref{t:varpar}.  The uncertainties in the derived values of 
$\delta_{\perp}$ have been estimated from monte carlo simulations
of the different datasets as described above: $\delta_{\perp}$ was 
calculated for each realization of the datasets, and the estimated 
uncertainties correspond to the standard deviation of the simulated 
values of $\delta_{\perp}$.

\begin{figure*}
\plotone{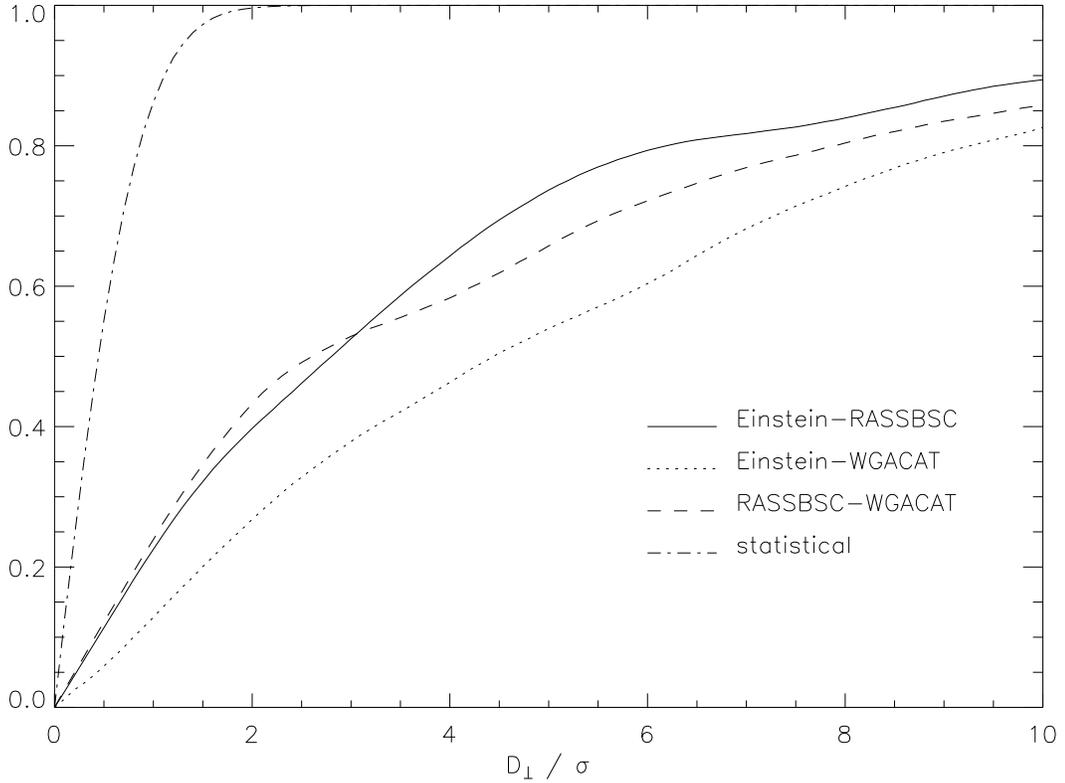}
\caption{Distribution of perpendicular deviates: the fraction of stars
in a given sample with deviates $\geq \frac{D_{\perp}}{\sigma}$ is shown.
In the presence of only statistical variations, the differential
distribution of $\frac{D_{\perp}}{\sigma}$ is a one-sided Gaussian,
indicated here by the dash-dotted line.  The presence of variability
in excess of statistical deviations in a chosen sample will result
in a larger fraction of the sample showing perpendicular deviates
at much larger values of $\sigma$, as is the case for the Einstein-RASSBSC
(solid stepped line), Einstein-WGACAT (dotted stepped line), and
RASSBSC-WGACAT (dashed stepped line) samples; note that $\sim$ 30, 45,
and 35\% of the stars in the respective samples show variability in
excess of $5 \sigma$.
\label{f:perpdev}}
\end{figure*}

In Figure~\ref{f:perpdev}, we show the cumulative distribution of
${D_{\perp}}/{\sigma_{tot}}$ computed for each paired dataset,
augmented by monte carlo simulations performed as described above in
order to illustrate the distributions more clearly.  The fraction of
stars in each sample with normalized perpendicular deviates $>
{D_{\perp}}/{\sigma_{tot}}$ are shown.  When the differences between
two samples may be attributed solely to statistical errors, the
differential distribution of normalized perpendicular deviates are
distributed as a one-sided Gaussian; this distribution is also
illustrated in Figure~\ref{f:perpdev}.  Any ``excess variability'' --
deviates larger than expected on purely statistical grounds or
systematic errors -- manifest themselves in the form of wider
distributions, i.e., with a larger fraction of stars in the sample
showing perpendicular deviates at larger values of
$D_{\perp}/\sigma_{tot}$.  Based on Figure~\ref{f:perpdev}, each of the
samples considered shows clear and unambiguous signatures of excess
variability.  Indeed, 30\% of the stars in the Einstein-RASSBSC sample,
45\% of those in the Einstein-WGACAT sample, and 35\% of those in the
RASSBSC-WGACAT sample show scatter attributable to non-statistical
variability at a level $>5 \sigma$.

We have also carried out a similar analysis on subsamples of the largest of
our three samples (Einstein-RASSBSC) in order to investigate whether or not
there are any trends in $D_{\perp}/\sigma_{tot}$ with spectral type or
luminosity class.  We find that the resulting distributions of
${D_{\perp}}/{\sigma_{tot}}$ are similar to the distribution obtained for the
full sample, indicating that in our data there is no significant evidence
for such systematic changes in observed scatter or variability in soft X-ray
emission.

One of the primary goals of this study is to look for evidence of
underlying variability with characteristic timescales of order a decade
or so, similar to that of the solar cycle.  In the case of the Sun,
such variability in soft X-rays is about an order of magnitude
(e.g., Pallavicini 1993, Hempelmann et al.\ 1996) or more (Kreplin
1970, Aschwanden 1994, Acton 1996).  If such a component of variability were
present in the stars of our active binary sample, we would expect the
two {\it Einstein-ROSAT} samples to exhibit a larger spread in
${D_{\perp}}/{\sigma_{tot}}$ than the RASSBSC-WGACAT sample, since the
{\it Einstein} and {\it ROSAT} respective observations span an interval
more comparable to the expected period of the long-term variability.  That
such a signature is not easily discernible may be partly attributed to
the generally larger errors associated with the {\it Einstein}
measurements of count rates---note that the perpendicular deviates
considered here are normalized relative to the estimated statistical
error.  Thus, in order to show a similar effect as the RASSBSC-WGACAT
sample, the {\it Einstein-ROSAT} samples must have a correspondingly
larger intrinsic non-statistical differences.  However, as we have
emphasized above, soft X-ray variability over the solar cycle amounts
to an order of magnitude or more, which is well beyond the statistical
uncertainties in the {\it Einstein-ROSAT} comparisons.  Therefore, if
any long-term, or cyclic, component of variability is present in the
stars of our active binary sample, then the amplitude of this
variability must be much less than in the solar case.  In the following
sections we discuss the implications of this result.

\subsection{\label{s:signify}Stochastic Variability}

The derived values of $\delta_{\perp}$ (see Table~\ref{t:varpar})
conclusively show that the data are inconsistent, at a very high
significance, with the hypothesis that there are only statistical
variations in count rates among the 3 datasets acquired at different
epochs: i.e., we unambiguously detect the existence of excess
variation among the samples.

The nature of this excess scatter is however not as well-determined.
We rule out instrumental effects as being the main cause of the
observed scatter since it is seen even in the RASSBSC-WGACAT
sample.
In the cases involving
the IPC, we note that even though the passbands and instrument
sensitivities of {\it Einstein} and {\it ROSAT} differ, for spectra
generated from thermal plasma at temperatures between $5 \times 10^6$ and
$\lesssim 10^7$ K, which are the likely coronal temperatures of the stars
being considered (e.g., Schmitt et al.\ 1990, Dempsey et al.\ 1993), these
differences are small (cf.\ Wood et al.\ 1995) and the maximum error we are
likely to make in the $\frac{PSPC}{IPC}$ count-ratio is $\sim 10\%$.
monte carlo simulations of the datasets including this type of error show
that its effect on the value of $\delta_{\perp}$ is to offset it by
$\sim 0.2$ and is hence negligible.  We therefore conclude that the origin
of the detected excess variations is intrinsic.

We now investigate the possibility that all of the observed excess
variation can be attributed to stochastic variability, and then
whether or not we can discern any differences in the magnitude of 
such variabilities between the different pairs of data (i.e., stars
common to [Einstein,RASSBSC], [Einstein,WGACAT], or [RASSBSC,WGACAT]).
This is 
not entirely straightforward because the three sets of observations 
were obtained under different conditions and with different
instrumentation and have different measurement uncertainties.  
To do this we 
first assume that the variability detected here may be 
parameterized by modeling it as
purely stochastic variability, relative to the estimated statistical
error.  We emphasize that we carry out this modeling only as a means
to explore the range of $\delta_{\perp}$, and that it is not our
intention to claim that intrinsic variability in active binaries indeed
follows this pattern.  We assume that the variability may be characterized
by the parameter $\beta = \frac{\Delta I}{\sigma_I}$, where 
here $\Delta I$ represents the effective change in soft X-ray emission
from one observation to the next; $\beta$ then represents
the ratio of the magnitude of the variation and the observed error in the
count rate.  Note that this assumption obviously underestimates the
magnitude of {\em cyclic} variability, but is adequate to summarize our
results given the absence of detailed time traces of photometric and
X-ray brightness of the stars in the sample.
For the parameter $\beta$ to be physically meaningful, the
estimated errors must be insensitive to distance effects -- i.e., the
expected variability must not be a function of our special location.  For
the stellar sample in question (Table~\ref{t:starlist}), we note that the
X-ray luminosity spans a range $\frac{max}{min}[{L_x}] \approx 17400$,
much greater than distance induced flux variations ($\frac{max}{min}[d^2]
\approx 4700$).  The spread in count rates is therefore much larger than
variations induced by stellar distance (and errors therein); the bias
introduced into the analysis due to farther sources being weaker and
thus naturally having larger relative errors is thus minimized, and the
adopted parameter is a reasonable quantity to use to describe the
samples.  Comparison of $\beta$ derived from different datasets is however
still subject to the problem of different datasets having different
relative errors, and we account for this later.

We derive the appropriate value of $\beta$ for each dataset pair as
follows.  Starting from an arbitrary sample of count rates (we used
RASSBSC because it is the largest sample) and an assumed value of
$\beta$, we generated using monte carlo simulations two new sets of
count rates for each point.  The new count rates were obtained by
sampling from two Gaussians, both with means equal to the original
count rate but with different standard deviations $\sigma_1 =
\sigma_{tot}$, the estimated statistical error (see
Equation~\ref{e:dperp}), and $\sigma_2 = \sqrt{1 + \beta^2 } \cdot
\sigma_{tot}$.  A $\delta_{\perp}$ was then derived for the new pair of
simulated datasets.  This process was repeated for different values of
$\beta$, resulting in predicted values of $\delta_{\perp}$ as a
function of $\beta$.  For each dataset pair, $\beta$ was then derived
by comparing this function with the observed $\delta_{\perp}$.  Note
that by definition of $\delta_{\perp}$ and $\beta$, this process is
insensitive to details of the original sample such as number of points,
sizes of individual errors, etc.

The results are listed in
Table~\ref{t:varpar}: All samples are characterized by non-statistical
relative variabilities $\beta > 10$, with the Einstein-RASSBSC sample
being the lowest as expected (due to the relatively large errors on the
count rates); and despite the significantly higher $\delta_{\perp}$ of
the Einstein-WGACAT sample relative to the RASSBSC-WGACAT sample, the
range of relative variabilities $\beta$ overlap with each other,
suggesting that long-term (potentially cyclic) variability is similar
in magnitude to short-term (potentially episodic) variability.

The derived relative variabilities may also be used to estimate an
``effective variability'', $\frac{\Delta I}{I} \sim {\beta}/{<SNR>}$,
where $<SNR>$ is an average measure of the signal-to-noise ratio.
Inclusion of this factor further minimizes the stellar-distance bias in
$\beta$.  We thus derive (see Table~\ref{t:varpar}) $\frac{\Delta I}{I}
= (0.29-0.36), (0.34-0.43), (0.41-0.52)$ for RASSBSC-WGACAT,
Einstein-WGACAT, and Einstein-RASSBSC respectively.  We note that the
observed RASS and WGA count rates for FF Aqr are sharply different
($\frac{RASS}{WGA} \sim 15$), and attribute this to a likely flare
event during the {\it ROSAT} All-Sky Survey.  This one star contributes
$\approx 10\%$ of the measured\footnote{We are potentially interested
in detecting long-term cyclic variability, and hence would be justified
in isolating the effects of such variability by eliminating other
contributors to the measured variability.  Note however that we do not
exclude FF Aqr from our analyses because we are unable to unambiguously
identify cyclic variability in the chosen samples.} $\frac{\Delta
I}{I}$.  The long-term samples have systematically larger values of
$\frac{\Delta I}{I}$, but are not significantly different given the
size of the error bars, the possible systematic errors (see above), and
the unsuitability of the adopted parameterization to characterize
cyclic variability (see \S\ref{s:cyclic}).  Note that the measured
``effective variability'' over ``short'' timescales (RASSBSC-WGACAT) is
similar to that found by Ambruster, Sciortino, \& Golub (1987) by
photon-arrival-time analysis of {\it Einstein} observations of selected
stars over timescales ranging from $\sim 10^2 - 10^3$ s.  A similar
result was also found by Pallavicini, Tagliaferri, \& Stella (1990) in
their analysis of {\it EXOSAT} data of flare stars: they detect
variability at a variety of timescales ($\sim 3^m - >100^m$) in half
the stars in their sample at strengths ranging from 15\%-50\%.  Thus it
might not be unreasonable to attribute most, and perhaps all, of the
causes of the observed RASSBSC-WGACAT variability to processes (e.g.,
flares, rotational modulation, active region evolution) that operate on
such relatively short timescales.

\begin{table}[ht]
\tablenum{3}
\begin{center}
\caption{\label{t:varpar} Measured Variability}
\begin{tabular}{c c c c}
\hfil & \hfil & \hfil & \hfil \\
\hline
\hfil & Einstein-RASSBSC & Einstein-WGACAT & RASSBSC-WGACAT \\
\hline
$\delta_{\perp}$
		& $4.38 \pm 0.07$
		& $6.62 \pm 0.11$
		& $5.33 \pm 0.09$
		\\

\hfil & \hfil & \hfil & \hfil \\
$\beta$
		& $12^{\ > 11}_{\ < 14}$
		& $20^{\ > 17}_{\ < 22}$
		& $15^{\ > 13}_{\ < 17}$
		\\

\hfil & \hfil & \hfil & \hfil \\
$\frac{\Delta I}{I}$
		& $0.46^{\ > 0.41}_{\ < 0.52}$
		& $0.38^{\ > 0.34}_{\ < 0.44}$
		& $0.32^{\ > 0.29}_{\ < 0.37}$
		\\

\hline
\end{tabular}
\end{center}
\end{table}

\subsection{\label{s:cyclic}Cyclic Variability}

Our modeling described above searches for stochastic (e.g., flaring)
variability, and is insensitive to potential systematic (e.g., cyclic)
variability in a sample where measurements of X-ray flux from
individual stars are uncorrelated.  However, if we compare the derived
magnitude of the variability $\frac{\Delta I}{I}$ in the short-term
($\sim 1$ yr) with the long-term ($\sim 10$ yr) samples, we do find
indications of a larger variation over longer timescales.  This result
is statistically inconclusive since the error-bars on the variability
indices ($\frac{\Delta I}{I}$) derived for the various samples overlap,
and further because of limitations imposed by the parameterization
itself.

What fraction of the above variability is due to stochastic causes, and
what fraction is due to the effects of periodic causes?  In order to
address this question, and thereby derive upper limits on the magnitude
of a cyclic component to the variability, we model the flux variations as
due entirely to a sinusoidal component combined with a constant base
emission.  We write the X-ray flux at an
arbitrary time $t$,
\begin{equation}
\label{e:cycle}
f_x(t) = A_{cyc} \sin\left(\frac{2 \pi t}{P_{cyc}} + \phi\right) +
A_{cyc} + f_{x_0} \,,
\end{equation}
where $A_{cyc}$ is the amplitude, $P_{cyc}$ is the period, and $\phi$ is the
phase of the cyclic component, and $f_{x_0}$ is a non-varying base
emission.  Note that $f_x (t) \geq 0$ for all $t$.  The strength of the
cyclic activity may be parameterized by the ratio of cyclic to base emission
fluxes,
\begin{equation}
\zeta = \frac{2 A_{cyc}}{f_{x_0}} \,.
\end{equation}
Conversely, if $f_{obs}$ is the observed flux at, say, $t=0$, 
and $\zeta'$ is an estimated fraction of the cyclic component,
\begin{equation}
A_{cyc} = \frac{\zeta' f_{obs}}{2 + \zeta' ( 1 + \sin\phi )} \,.
\end{equation}

\begin{figure*}
\plotone{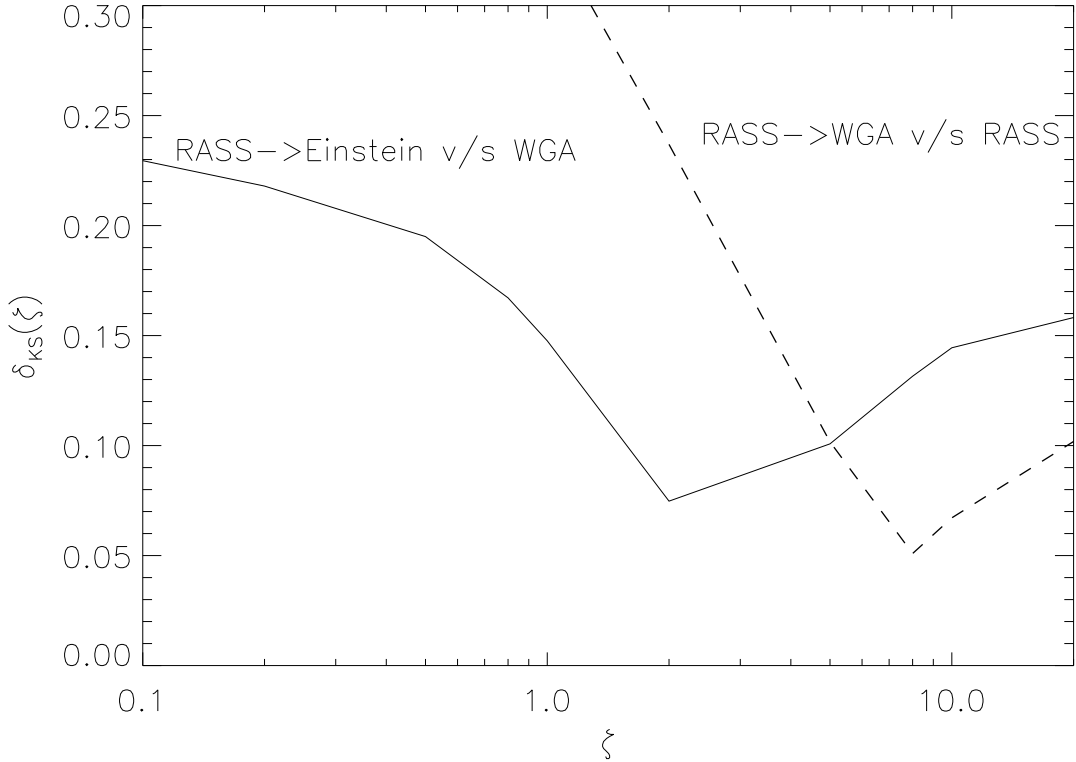}
\caption{Magnitude of cyclic variability: The value of intrinsic cyclic
amplitude that the selected set of stars must have in order that this
variability is detectable.  The maximal distance $\delta_{KS}(\zeta)$
between the cumulative distributions of $\delta_{\perp}$ for the modeled
dataset and the reference dataset is shown for various values of the
amplitude of cyclic variability for comparisons between observations 
separated by ``long'' (solid) and ``short'' (dashed) time intervals.
The location of the minimum indicates the best model value that fits
the data.  The magnitude of cyclic variability present in the long-term
(solid line) sample is $\lesssim 2$. Note that the error in
$\delta_{KS}(\zeta) \sim 0.03$, and hence a better estimate for
the amplitude of cyclic variability is $\zeta \lesssim 4$.
In the case of the short-term sample (dashed line), $\zeta \sim 10$,
which predicts variations that are much larger than observed over the
long-term (cf.\ Figures \ref{f:einrass}-\ref{f:rasswga}).  This latter
result indicates that cycles alone {\em cannot} be responsible for the
observed variability (see text).
\label{f:cycvar}}
\end{figure*}

We then estimate the maximum value that $\zeta$ can have for our
adopted sample of stars, using a technique similar to that used to
measure the strength of stochastic variability (\S\ref{s:correl}).  In
order to minimize the effects of short-term variability on estimates of
cyclic variability, we model the cyclic variability {\it starting from}
a paired dataset (say $A$ and $B$).  The modeling involves obtaining
monte carlo transpositions of the count rates of one the samples (say
$A$) to the epoch of a different sample (say $C$).  This transposition
is carried out for a fixed value for the strength of the cyclic
component (i.e., $\zeta = const.$), and using values of $P_{cyc}$
randomly sampled from a log-normal distribution with a mean
corresponding to 10 yr and $1 \sigma$ range corresponding to 4-25 yr
(this range is a rough approximation of the results tabulated by
Baliunas et al.\ 1995, based on the Mt.Wilson Ca\,II H+K monitoring
program; see their Figure 3), at randomly selected phases, and
including the effects of the error bars as in \S\ref{s:correl} .  A
distribution of $D_{\perp}/\sigma_{tot}$ is obtained as before for the
paired datasets of the model ([$A \rightarrow C$]($\zeta$)) and the
original dataset ($B$) -- in other words, for the {\it simulated} pair
$B-C$ -- and is compared with the distribution derived from the
reference dataset of paired samples (here, the {\it observed} pair
$B-C$).  The value of $\zeta$ that minimizes the difference between the
modeled and reference distributions of $D_{\perp}/\sigma_{tot}$
indicates the level of cyclic variability required in order to account
for the difference between the two pairs of datasets, {\it if} the
entire difference is to be attributed to cyclic variability.  The
difference between the distributions is simply parameterized by the
Kolmogorov-Smirnoff test statistic of the maximal distance
$\delta_{KS}(\zeta)$ between the cumulative distributions of
$D_{\perp}/\sigma_{tot}$.  Note that we do not attempt to derive a
probability value for this statistic, partly because doing so
implicitly assumes that the model adopted for transposing the count
rates (Equation~\ref{e:cycle}) is correct, and partly because the
modeling is carried out via monte carlo simulations.  Thus, we set an
upper limit on the amplitude of the cyclic variability observable in
the paired datasets $B$ and $C$.  Since we have three datasets, there
are many combinations possible since each of RASSBSC, Einstein and
WGACAT can correspond to all or any of $A$, $B$ and $C$.  The curves
illustrating the Kolmogorov-Smirnoff test statistic as a function of
$\zeta$ for two cases are illustrated in Figure~\ref{f:cycvar}.

The derived limit does depend on the sampling distribution adopted for
the cyclic periods (see above): if longer periods are preferentially
sampled (e.g., by increasing the width of the distribution, or by shifting
it to longer periods), the upper limit on $\zeta$ decreases -- i.e.,
smaller cyclic amplitudes are sufficient to account for the entire change
in the count rates of the samples studied -- and vice versa.  This trend
is however weak when compared to the adopted distribution of $P_{cyc}$.

In order to investigate the cyclic component with greatest
sensitivity, the observations obtained with longer time intervals 
between them need to be compared.  In this case, our datasets $A$, $B$
and $C$ corresponded to RASSBSC, WGACAT and Einstein, respectively.
The function $\delta_{KS}(\zeta)$ is illustrated for this case in
Figure~\ref{f:cycvar}:
considering paired samples of [RASSBSC$\rightarrow$Einstein]($\zeta$)
and WGACAT (in other words, simulated Einstein-WGACAT) with a directly
observed sample of Einstein-WGACAT, we find that the
actual cyclic component must have a strength $\zeta \lesssim 4$ -- i.e., not
more than 80\% of the X-ray emission may be in the cyclic component
even if all of the long-term variability is ascribed to the effects
of activity cycles.
It is worth noting here that we implicitly include the effects of
short-term variability by using the RASSBSC-WGACAT sample as the
initial distribution for modeling.  The cyclic component is thus imposed
on top of any variations that may exist due to flaring, and the effects
of the latter are thus minimized.\footnote{The
structure of our chosen datasets do not allow us to distinguish between
``flaring'' and ``quiescent'' (which may arise due to active region
evolution, rotational modulation, etc.) variability, a distinction made
by K\"{u}rster et al.\ (1997) using periodogram analysis of the light curve
of AB Dor; the ``short-term'' variability we refer to includes both types.
}

Also worthy of investigation is the magnitude of cyclic variability
that might be required to explain {\em all} the additional
(non-statistical) scatter between the RASSBSC and WGACAT samples---ie
the assumption that there is no short-term or stochastic variability.
For example, starting with a sample of RASSBSC stars alone,
transposing them to the approximate WGACAT epoch for various values of
$\zeta$, and considering the resultant paired sample of
[RASSBSC$\rightarrow$WGACAT]($\zeta$)
and RASSBSC, we find that only at large values of the cyclic component
($\zeta \sim 10$) does the distribution match that derived from
RASSBSC-WGACAT (Figure~\ref{f:cycvar}).  However, such a large
amplitude for cyclic variability would result in a much larger spread
in count rates than is observed in our comparisons of Einstein vs.\
ROSAT datasets (Figures~\ref{f:einrass} and \ref{f:einwga}).  This is
as one would expect in the presence of significant short-term
stochastic variability: indeed, it is telling us that the observed
variability {\em cannot} only be of a cyclic nature.  This constraint
is only possible using observations obtained at three different 
epochs, as we have here.


It is interesting to compare our upper limit for $\zeta$ for this sample
of active binary stars to the observed cyclic component of solar
coronal activity.  In the case of the Sun, data obtained by {\it Yohkoh}
have illustrated the very large contrast in soft X-rays between the
Sun at solar minimum, essentially devoid of active regions, and at
solar maximum when several large active regions are generally present
on the visible hemisphere at any one time (e.g., Hara 1996, Acton 1996).
The observed change in soft X-ray flux from solar minimum to maximum
amounts to at least an order of magnitude (Aschwanden 1994, Acton 1996)
and likely much larger (Kreplin 1970), so that a solar value for our
parameter describing the cyclic activity component is
$\zeta_{\odot} \gtrsim 10$.

Fleming et al.\ (1995) have carried out an analysis of X-ray variability
using an X-ray selected sample, viz., RASS flux measurements of EMSS stars.
They find that relative to F stars, 24\% of Solar-type stars, 49\% of
dMe stars, and 19\% of RS\,CVn and W\,UMa stars show a significant {\em
decrease} in emission, while 12\%, 10\%, and 48\% respectively of the above
types show significant {\em increases}.  The larger apparent decreases for
normal stars may be ascribed to the bias inherent in X-ray selected stellar
samples.  In contrast, the apparent increase in the number of active binary
stars showing increased emission levels cannot be due to such a bias.  We
are however unable to confirm this effect using a different, and larger,
sample of stars that include EOSCAT and Slew observations.  Indeed, for
unbiased, uncorrelated, samples it is difficult to envision a physical
mechanism that would cause this apparent increase found by them, and we
therefore attribute it to accidental sample selection effects (as Fleming
et al.\ also do) and to possible variations in flux calibration (see Figure
3 of Fleming et al).

Comparison of X-ray fluxes between {\it Einstein} and {\it ROSAT} epochs
using statistically complete samples of field stars (Schmitt et al.\ 1995)
shows that there is little evidence for systematic changes in the mean 
X-ray emission levels of stars in excess of factors of 2 on timescales
of 10 yr.  Schmitt et al.\ point out that this result is valid especially
for active flare stars, but caution that the apparently larger spread
in X-ray emission observed for fainter stars could be an artifact of the
small number of such stars present in the sample.

Attempts to constrain long-term variability in X-ray emission from
stars in the Pleiades (Schmitt et al.\ 1993, Gagn\'{e} et al.\ 1995,
Micela et al.\ 1996) and the Hyades (Stern et al.\ 1995) clusters have
also been inconclusive.  Schmitt et al.\ (1993) compare Pleiades stars
detected in the {\it ROSAT} All-Sky Survey with previous {\it Einstein}
observations and find numerous instances of strong variability by
factors of an order of magnitude that are unlikely to be due to
rotational modulation, measurement or calibration errors, or flaring
activity; they conclude that cyclic activity must be the cause of such
large variations.  In contrast, Gagn\'{e} et al.\ (1995) and Micela et
al.\ (1996) find at best a marginal increase in long-term variability
compared to short-term variability in their analyses of pointed data:
the latter find that 15\% of the stars show variability by factors $>2$
over $\sim 10$ years and 10\% over $\frac{1}{2}$ year; the former find
40\% of the stars show significant variability over $10-11$ years
compared to 25\% over $\frac{1}{2}-\frac{3}{2}$ years, but that the
difference could be attributed to a bias resulting from the increased
sensitivity of {\it ROSAT}.  Stern et al.\ (1995) conclude from their
comparison of {\it Einstein} and {\it ROSAT} data of the Hyades that
the majority of the stars show long-term variability of less than a
factor of 2, and that there is no evidence of strong cyclic activity.
As noted above, this amplitude of variability is similar to what one
would expect on short timescales of $\lesssim 10^3$ s based on the
time-series analyses of {\it Einstein} observations of various stars
(Ambruster et al.\ 1987).

In this context, it must be noted that a long-term monitoring program of
the young active K star AB Dor has been carried out with {\it ROSAT} by
K\"{u}rster et al.\ (1997): they find that the X-ray flux is variable on
short time-scales ranging from minutes to weeks, but shows no long-term
trend indicative of cyclic activity over the $5\frac{1}{2}$ years of the
program.  The lack of a detection of cyclic variability in this star
of course does not rule out its presence in other active stars, and indeed
may even manifest itself in AB Dor itself at much longer timescales.

Micela \& Marino (1998) have compared the changes in X-ray emission
in field dM stars observed with {\it ROSAT} over timescales of days
to months, and have compared that data with similar data for the Sun
by constructing maximum-likelihood distribution functions of the flux
variations.  They conclude that variability is present at all timescales
they have considered, but do not distinguish between long- and short-term
or stochastic and cyclic variability.  They have also applied their method
to compare Pleiades dM star data from {\it Einstein}/IPC, {\it ROSAT}/PSPC,
and {\it ROSAT}/HRI to Solar data and reach the same conclusion.

The problem of whether or not F-K main sequence stars in general are
similar to the Sun in their trends of X-ray emission variations with
magnetic cycles has been studied recently by Hempelmann et al.\ (1996).
These authors attempted to circumvent the sparse X-ray observations of
any one single star through a statistical study of a group of stars
for which long-term Mt.Wilson Ca\,II H+K monitoring observations are
available, and for which distinct cyclic activity behavior was detected
(e.g., Baliunas et al.\ 1995 and references therein).  They found that
the {\em deviation} of X-ray surface fluxes $F_{\rm x}$ (derived from
{\it ROSAT} all-sky survey and pointed phased observations) from the mean
relation between $F_{\rm x}$ and Rossby number, $R_o$, is correlated with
activity cycle phase as indicated by the Mt.Wilson Ca\,II H+K $S$-index.

In addition to stars with cyclic Ca\,II H+K emission, the Mt.Wilson
monitoring program also revealed stars with less regular and more
chaotic variability (Baliunas et al.\ 1995).  Hempelmann et al.\ (1995)
showed that the ``regular'' (cyclic) and ``irregular'' stars are strongly
anti-correlated with the Rossby number, $R_o$; the X-ray fluxes of the
latter group clearly show them to comprise the most active stars with the
highest surface fluxes.  Hempelmann et al.\ interpreted these results
in terms of a transition from a non-linear to a linear dynamo going from
the irregular to the regular stars.  This view is supported by non-linear
modeling of stellar dynamos (e.g.\ Tobias, Weiss, \& Kirk 1995, Knobloch
\& Landsberg 1996, and references therein) which show that as stellar
rotation period is decreased, an initially steady system begins to exhibit
quasi-periodicity or maunder-minimum type aperiodicity; chaotic behavior
is a natural consequence of such models.

%
Alternately,
Drake et al.\ (1996) argued that the observational evidence indicating
both active stars and fully-convective M-dwarfs---the latter supposedly
being unable to support a solar-like $\alpha\,\omega$ dynamo---do not
appear to show strong cyclic behavior provided empirical support for
the qualitative theoretical framework outlined by Weiss (1993; see also
Weiss 1996) in which the magnetic activity on active stars and low-mass
fully-convective stars is predominantly maintained by a turbulent or
distributed dynamo (e.g., Durney et al.\ 1993, Weiss 1993, Rosner et al.\
1995).  Stern et al.\ (1995; see also Stern 1998), in their comparison
of {\it Einstein} and {\it ROSAT} observations of Hyades dG stars, made
similar speculations.  Small-scale magnetic fields generated by turbulent
compressible convection within the entire convective zone (e.g., Nordlund
et al.\ 1992, Durney et al.\ 1993) would result in different functional
dependences of activity indicators with stellar parameters, and in
particular, because of its disordered spatial and temporal nature, would
not be expected to exhibit activity cycles (Cattaneo 1997).

In the above context, the general lack of well-defined activity cycles in
the Ca\,II H and K emission of the most active stars is especially
interesting in light of the very recent finding based on radio observations
of a magnetic cycle on the rapidly rotating RS\,CVn binary UX\,Ari (G0\,V +
K0\,IV; Period=6.44 day), with an apparent polarity reversal every 25.5
{\em days} (Massi et al.\ 1998).  If this surprisingly short period
were analogous to the solar 22 year cycle, it would appear to offer a
promising explanation for the lack of obvious cyclic behavior on
timescales of years.  However, the situation regarding cycles on
RS\,CVn and other active stars is not quite so clear.  Several authors
have found evidence for cyclic behavior with periods comparable to
that of the solar cycle on very active stars.  For example, cycles of
11.1 yr for $\lambda$ And, 8.5 yr for $\sigma$\,Gem, 11 yr for II Peg, and
16 yr for V711\,Tau were inferred by Henry et al.\ (1995) from mean
brightness changes derived from up to 19 years of photoelectric
photometry. Dorren \& Guinan (1994) find evidence for an activity
cycle with a period of about 12 years on the rapidly rotating
solar-like G dwarf EK\,Dra (HD\,129333; rotation period $\sim 2.8$
days).  Rodon\`o et al.\ (1995) also find a periodic variation in the
spot coverage of RS\,CVn with a period of about 20 years, while Lanza
et al.\ (1998) find similar evidence for a 17 year activity cycle in
the secondary of AR\,Lac.  Alternatively, others have detected photometric
variability, but failed to find firm evidence for cyclic behavior in
numerous binaries (eg.\ Strassmeier et al.\ 1994 in the RS\,CVn star
HR\,7275; Ol\'ah et al.\ 1997 in the case of HK\,Lac; Cutispoto 1993 in
the case of IL\,Hya, LQ\,Hya, V829\,Cen, V851\,Cen, V841\,Cen, GX\,Lib,
V343\,Nor, and V824\,Ara).

Regardless of the observed periods of activity cycles of active
binaries based on optical observations of modulation in, presumably,
magnetically-related spot coverage, it is clear based on the results of
our and earlier analyses that magnetic activity manifest in coronal
X-ray emission is at best only weakly modulated by any long-term
magnetic cycles present.  This contrasts with the solar case in which
coronal activity is very strongly dependent on the solar magnetic
cycle.  While we cannot rule out the existence of a multi-period
cyclic dynamo, we suggest based on the relatively small difference
between the short-term and long-term variabilities that this situation
reinforces the earlier conclusions and conjectures of Stern et al.\
(1995), Drake et al.\ (1996) and Stern (1998) that a turbulent or
distributed dynamo dominates the magnetic activity of the more active
stars.

\section{\label{s:summary}Summary}

In order to determine the characteristics of the variability of X-ray
emission on active binary systems, we have carried out a statistical
comparison of the X-ray count rates observed at different epochs.
From the list of active chromosphere stars cataloged by Strassmeier
et al.\ (1993) we have extracted subsamples which were detected in
the following surveys: {\it Einstein}/IPC EOSCAT, EMSS, and Slew;
{\it ROSAT} All-Sky Survey (RASSBSC); {\it ROSAT} archival pointed
dataset (WGACAT).  Our study differs from and improves upon earlier
comparisons of {\it Einstein} and {\it ROSAT} observations of late-type
stars in that the analysis of both RASSBSC and WGACAT observations enables
us, at least in principle, to distinguish between ``short'' and long-term
components of variability.

Assuming that the emission from separate stars are uncorrelated, we
compute a measure of the relative departure from equality ($\delta_{\perp}$,
Equation~\ref{e:dperp}) for each combination of the above samples.  We
show that the values of $\delta_{\perp}$ thus derived are inconsistent
with each other, i.e., there is evidence for non-statistical variations
in the observed count rates of the sample stars in the different epochs.

We model the detected variability as stochastic variability, and
conclude that the ``effective variability'' (an average value of the
fractional variation in the observed count rate) is apparently the
lowest for the samples separated by the shortest timescales
(RASSBSC-WGACAT:  $\frac{\Delta I}{I} = 0.32_{\ 0.29}^{\ 0.36}$), and
appears to be systematically larger for the samples separated by
longer timescales (Einstein-RASSBSC:
$\frac{\Delta I}{I} = 0.46_{\ 0.41}^{\ 0.52}$;
Einstein-WGACAT: $\frac{\Delta I}{I} = 0.38_{\ 0.34}^{\ 0.43}$).
This suggests the existence of a long-term component to the variability,
but the evidence for such a component is marginal.  If such a component
exists, it could be due to stellar activity cycles strongly modified
by X-ray emission arising due to relatively unmodulated small-scale fields
generated by turbulent dynamos in the convective zone.

We model the long-term component as a sinusoidal cyclic variation atop
a constant base emission, and constrain its strength by comparing
simulated distributions of perpendicular deviations with observed
distributions.  We find that such a cyclic component, if it exists,
may at most be 4 times as strong as a constant, base emission.  This
contrasts with the Solar case, where cyclic activity causes an
increase in the soft X-ray emission by factors $\gtrsim 10$ at
activity maximum relative to the flux at activity minimum.  We note
earlier conclusions that the nature of coronal activity on active
stars fits the scenario whereby the generation of magnetic fields
whose dissipation is observed in the form of coronal heating and
subsequent radiative loss is dominated by a turbulent or distributed
dynamo, rather than by a solar-like $\alpha\,\omega$ large-scale field
dynamo.  This scenario is essentially the same as that suggested by,
e.g., Weiss (1993, 1996), based on qualitative theoretical arguments.
Comparisons of past and future observations of stellar coronal
emission at different epochs, such as those analyzed here, for larger
samples of stars with different activity levels and different spectral
types, will be invaluable in distinguishing between different
dynamo models.

\acknowledgments

We would like to thank Alanna Connors, Steve Saar, Brian Wood, Frank
Primini, and the referee for useful comments.
This research has made use of the SIMBAD database, operated at CDS,
Strasbourg, France.  VK was supported by NASA grants NAG5-3173,
NAG5-3189, NAG5-3195, NAG5-3196, NAG5-3831, NAG5-6755 and NAG5-7226
during the course of this research.  JJD was supported by the AXAF
Science Center NASA contract NAS8-39073.

\end{document}